\documentclass[multphys,vecphys]{svmult}

\usepackage{makeidx}     
\usepackage{graphicx}    
\usepackage{multicol}    
\makeindex             

\begin{document}

\title*{The Radio Afterglows of Gamma-Ray Bursts}
\author{Dale A. Frail} 
\institute{National Radio Astronomy Observatory, Socorro, NM 87801 USA}
\maketitle

\section{Introduction}\label{sec:intro}

Our understanding of the gamma-ray bursts (GRBs) has advanced rapidly
since the discovery of long-lived ``afterglow'' emission from these
events. Radio afterglow studies have become an integral part of this
field, providing complementary and sometimes unique diagnostics on GRB
explosions, their progenitors, and their environments. The reason for
this is that the radio part of the spectrum is phenomenologically
rich.  This can be illustrated simply by calculating the brightness
temperature ($T_b\propto {\rm F}_\nu/(\theta_s\,\nu)^2$) for a 1 mJy
centimeter wavelength source at cosmological distances ($\sim 10^{28}$
cm), expanding with $V_{exp}\leq c$ one week after the burst. Since
the derived $T_b\sim 10^{13}$ K is well in excess of the $T_{IC}\sim
10^{11}-10^{12}$ K limit imposed by inverse Compton cooling, it
follows, independent of any specific afterglow model, that the radio
emission must originate from a compact, synchrotron-emitting source
that is expanding superluminally (i.e. $T_b\sim\Gamma\times T_{IC}$,
$\Gamma>>1$). Likewise, since the brightness temperature cannot exceed
the mean kinetic energy of the electrons, the emission is expected to
be self-absorbed at longer wavelengths \cite{kp97}.  Finally, strong
modulation of the centimeter signal is expected on timescales of hours
and days because the angular size $\theta_s$ of this superluminal
source is comparable to the Fresnel angle of the turbulent ionized gas
in our Galaxy \cite{goo97}. Synchrotron self-absorption, interstellar
scintillation, forward shocks, reverse shocks, jet-breaks,
non-relativistic transitions and obscured star formation are among
the phenomena routinely observed.

This short review is divided into two parts. The first section
(\S\ref{sec:obs}) is a summary of the current search strategies and
the main observational properties of radio afterglows. In the second
section (\S\ref{sec:cont}) we highlight the key scientific
contributions made by radio observations, either alone or as part of
panchromatic studies. By necessity we will restrict this brief review
to long-duration GRBs, although radio afterglows have also been
detected toward the newly classified X-ray flashes, and searches have
been carried out toward short bursts \cite{hbc+02}.

\section{Detection Statistics and Observational Properties}\label{sec:obs}

The search for a radio afterglow is initiated either by a satellite
localization of the burst, or by the detection of the X-ray or optical
afterglow.  The current search strategy has been to use the Very Large
Array (VLA)\footnotemark\footnotetext{The NRAO is a facility of the
  National Science Foundation operated under cooperative agreement by
  Associated Universities, Inc.} or the Australia Telescope Compact
Array (ATCA; for declinations,
$\delta<-40^\circ$)\footnotemark\footnotetext{The Australia Telescope
  is funded by the Commonwealth of Australia for operation as a
  National Facility managed by CSIRO.} at 5 GHz or 8.5 GHz.  These
frequencies were chosen as a compromise between the need to image the
typical error box size of 30-100 arcmin$^2$, while having the
requisite sensitivity to detect afterglows at sub-milliJansky levels.
At lower frequencies the afterglow is attenuated by synchrotron
self-absorption ($f_\nu\propto\nu^2$), while at higher frequencies the
field-of-view is proportionally smaller (FOV$\propto\nu^2$). For
typical integration times (10 min at the VLA, and 240 min at the ATCA)
the $rms$ (receiver) noise is 30-50 $\mu$Jy.  Follow-up observations
of detected afterglows were carried out by a network of radio
facilities at centimeter, millimeter and submillimeter wavelengths
\cite{fkw+00}.

\begin{figure}
\centering
\includegraphics[height=4.9cm]{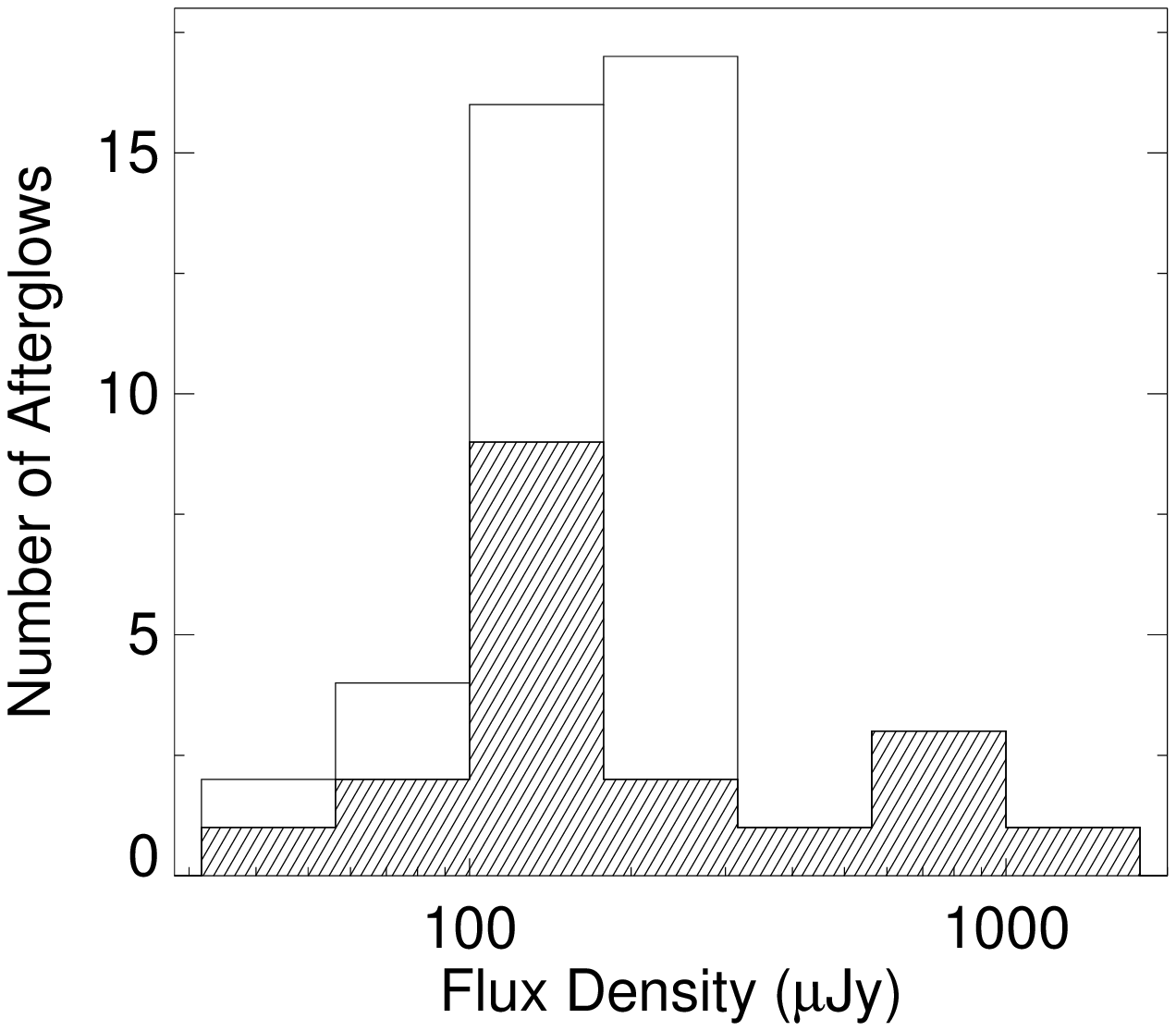}
\includegraphics[height=4.85cm]{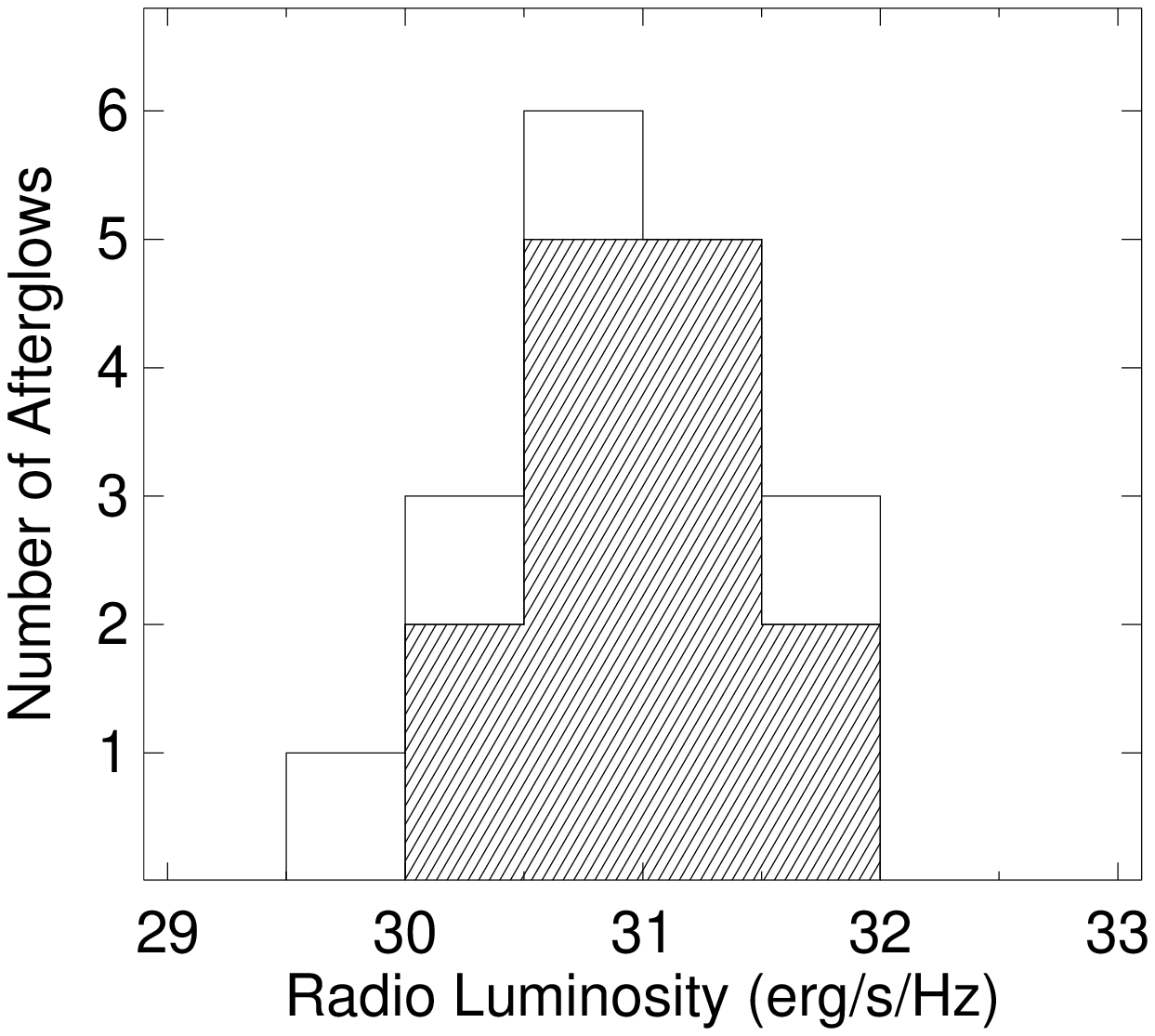}
\caption{({\it Left}) Histogram distribution of flux densities 
  (or upper limits) at 8.5 GHz for a complete sample of bursts. The
  hatched histogram shows the distribution for the detections only.
  ({\it Right}) Histograms of radio luminosity from the same sample
  but restricted to the subset of bursts with known redshifts.  The
  hatched histogram shows the distribution for bursts with detected
  radio afterglows only.
\label{fig:flux_lum}}
\end{figure}

In the five year period beginning in 1997 and ending in 2001
approximately 1500 radio flux density measurements (or upper limits)
were made toward 75 bursts \cite{fkbw03}. From these 75 GRBs, there
are a total of 32/36 successful X-ray searches, 27/70 successful
optical searches, and 25/75 successful searches. These afterglow
search statistics illustrate a well-known result, namely that the
detection probability for X-ray afterglows is near unity, while for
optical afterglows and radio afterglows it is 40\% and 33\%,
respectively. The origin of these optically ``dark bursts'' could
either be due to intrinsic effects (i.e. inadequate search due to
rapid evolution of the afterglow and/or an under-energetic GRB)
\cite{fjg+01,bkb+02}, or an extrinsic effect (i.e. extinction of the
optical flux caused by circumburst dust or by the intergalactic
medium) \cite{pfg+02,dfk+01}.

To accurately derive the fraction of ``radio quiet'' bursts it is
necessary to incorporate both detections and upper limits in a
statistically sound manner. This has been done in
Fig.~\ref{fig:flux_lum} where flux density distribution at 8.5 GHz is
shown for a sample of 44 GRBs, toward which measurements or upper
limits have been made between 5 and 10 days after a burst. The time
since the burst is an important variable since radio light curves do
not exhibit the simple power-law decays seen in X-ray and optical
afterglows, but rise to a peak on average about one week after the
burst and decay on timescales of a month.  The mean of the 19 {\it
  detections} in Fig.~\ref{fig:flux_lum} is 315$\pm$82 $\mu$Jy. Adding
in the non-detections, and using the Kaplan-Meier estimator
\cite{fn85} shifts this to 186$\pm$40 $\mu$Jy.  Approximately 50\% of
all bursts have radio afterglows at 8.5 GHz above 110 $\mu$Jy, while
fewer than 10\% exceed 500 $\mu$Jy. The relatively small range of peak
flux densities in Fig.~\ref{fig:flux_lum} suggests that the fraction
of ``radio quiet'' bursts is largely determined by instrumental
sensitivities. With the arcsecond localizations provided by the {\it
  Swift} satellite (launch in 2004) it will be possible to routinely
detect all afterglows with centimeter radio emission above 100
$\mu$Jy.  Increasing the fraction of detected radio afterglows
significantly above 50\% will require the sensitivity improvements
provided by the {\it Expanded Very Large
  Array}\footnotemark\footnotetext{http://www.aoc.nrao.edu/evla/}
(complete in 2010).

From this sample of peak flux densities we also derive the peak
spectral radio luminosity in Fig.~\ref{fig:flux_lum} given by
$L_\nu=4\pi F_\nu\,d_L^2\,(1+z)^{1+\beta-\alpha}$, where
$F_\nu\propto{t^\alpha}\nu^{\beta}$ and $\alpha=1/2$ and $\beta=1/3$
has been assumed, corresponding to an optically thin, rising light
curve. The GRB redshifts lie in the range between $z=$0.36 to $z$=4.5.
The peak of the distribution is centered on 10$^{31}$ erg s$^{-1}$
Hz$^{-1}$ and is similar to low-luminosity FRI radio galaxies like
M87. More interestingly, a comparison between this GRB sample and a
sample of Type Ib/c supernovae \cite{bkfs03} shows that the later is
four orders of magnitude less luminous. Since radio emission is
sensitive to the relativistic energy content of the shock, independent
of the initial geometry of the explosion, this has been used to argue
that the majority ($<$97\%) of nearby Type Ib/c supernovae do not
produce a GRB-like event, such as that seen toward SN\,1998bw
\cite{kfw+98}.

\section{Phenomenology and Interpretation}\label{sec:cont}

\begin{figure}
\centering
\includegraphics[height=6.85cm]{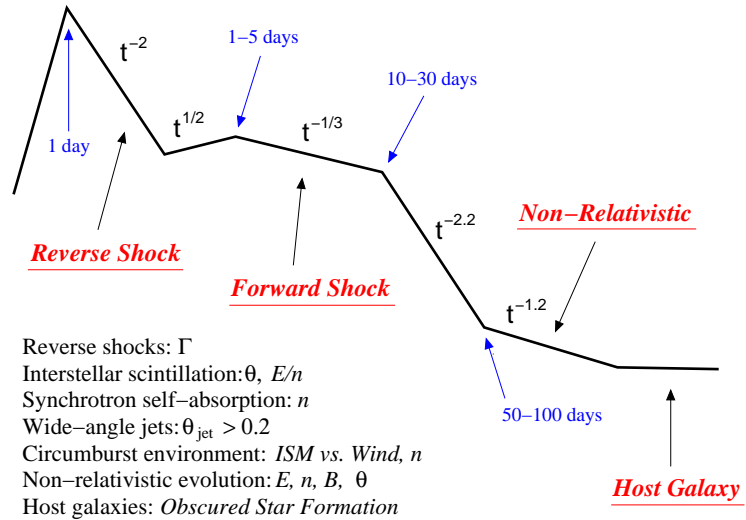}
\caption{A schematic radio afterglow light curve. Timescales and
  scalings for the temporal evolution are indicated.  The list
  summarizes aspects of the flux evolution which are unique to the
  radio bands (Lorentz factor, $\Gamma$; source size, $\theta$;
  energy, $E$; density, $n$; jet opening angle, $\theta_{\rm jet}$;
  density profile; magnetic field strength, $B$; and obscured star
  formation rate).
\label{fig:cartoon}}
\end{figure}

In this section we will follow the evolution a GRB and its radio
afterglow depicted schematically in Fig.~\ref{fig:cartoon}. The
observations span four orders of magnitude in time (0.1-1000 days) and
three orders of magnitude in frequency (0.8-660 GHz), so it should be
no surprise that radio light curves exhibit a rich phenomenology. To
interpret these observations we will rely on the highly successful
``standard fireball model'' \cite{mes02}. In this model there is an
impulsive release of kinetic energy ($\sim$10$^{51}$ erg) from the GRB
event which drives an ultra-relativistic outflow into the surrounding
medium whose hydrodynamical evolution is governed by the kinetic
energy released, the density structure of the circumburst medium and
the geometry of the outflow. Synchrotron emission is produced by this
relativistic shock which accelerates electrons to a power-law
distribution.  It is through the study of temporal (and spectral)
evolution of afterglow light curves that we can gain insight into the
physical conditions of the shock and the central engine that produced
it.

Despite response times as short as 2 hrs, centimeter searches
(\S\ref{sec:obs}) are rarely successful until a day or more after a
burst. Broadband afterglow spectra show that centimeter emission is
attenuated as a result of synchrotron self-absorption \cite{gbb+00}.
Typical observed values for the self-absorption frequency $\nu_a$ are
5-10 GHz. It is interesting to note that the flux density below
$\nu_a$ has the form F$_\nu\propto\nu^{2}$, not the 5/2 spectral slope
usually seen toward most radio sources. This is because the
relativistic shock accelerates electrons to a power-law distribution
(with energy index $p$ given by N($\gamma_e)\propto\gamma_e^{-p}$)
above a minimum energy $\gamma_m$, which initially radiate their
energy most of their energy at $\nu_m>>\nu_a$. The flux below $\nu_a$
depends only the angular size of the source and the fraction of the
shock energy that goes into accelerating electrons \cite{kp97}, and
thus it is a useful diagnostic of the ratio of the energy of the shock
and the density of the circumburst medium ($E/n$).

As this optically thick radio source expands, a monotonic rise in the
flux would be expected. It was therefore a considerable surprise when
early observations of GRB\,970508 showed erratic, short term
($\sim$ hrs) and narrow band ($\sim$GHz) fluctuations in the
centimeter emission \cite{fkn+97}. The origin of these variations
\cite{goo97} was traced to the scattering of the radio emission, owing
to the small angular size of the fireball, as it propagates through
the turbulent ionized gas of our Galaxy. This is a large and complex
subject \cite{rick90,gfs+03}, but for the purpose of this review it is
sufficient to note that for typical lines of sight the modulation of
the flux densities is near a maximum at frequencies near 5-10 GHz.
Coincidently, this is the same frequency range where $\nu_a$ typically
lies and where the majority of radio observations are being made.
While interstellar scintillation adds a certain degree of complexity
to interpreting afterglow light curves, it also allows us to use the
Galaxy as a large lens to effectively resolve the fireball. The
observed ``quenching'' of diffractive scintillation from GRB\,970708
four weeks after the burst \cite{wkf98,fwk00} lead to estimate of the
angular size, demonstrating superluminal expansion and providing an
early confirmation of the fireball model.

In many instances \cite{kfs+99,fbg+00,hys+01,yfh+02,bsfk03} bright,
short-lived radio ``flares'' are detected at early times ($t<3$ d).
The emission is much brighter than expected from a backward
extrapolation of the light curve, and the level of fluctuation is too
great to be accounted for by interstellar scintillation. One of the
best-known examples is the radio flare of GRB\,990123 \cite{kfs+99},
which was accompanied by a 9th magnitude optical flash
\cite{abb+99}.  This prompt optical and radio emission is thought to
be produced in a strong reverse shock which adiabatically cools as it
expands back through the relativistic ejecta \cite{sp99b}. The
strength and lifetime of this reverse shock emission is sensitive to
the initial Lorentz factor, $\Gamma_\circ$, of the shock and the
density structure of the circumburst medium \cite{sr02,bsfk03}. To
properly constrain these values requires that the peak of the emission
be measured. This is difficult to do with optical observations, which
require a response time on the order of the burst duration, while
radio observations require a response time of only 12-48 hrs.

On a timescale of days to weeks after the burst, the subsequent
evolution of the radio afterglow (Fig.~\ref{fig:cartoon}) can be
described by a slow rise to maximum, followed by a power-law decay.
The radio peak is often accompanied by a sharp break in the optical
(or X-ray) light curves \cite{hbf+99,bsf+00}. The most commonly
accepted (but not universal) explanation for these achromatic breaks
is that GRB outflows are collimated. The change in spectral slope,
$\alpha$, where $F_\nu\propto t^\alpha \nu^\beta$, occurs when the
$\Gamma$ of the shock drops below $\theta_j^{-1}$, the inverse opening
angle of the jet \cite{rho99,sph99}. Since the radio emission at
$\nu_R$ initially lies below the synchrotron peak frequency $\nu_m$
the jet break signature is distinctly different than that at optical
and X-ray wavelengths. Prior to the passage of $\nu_m$ the jet break
is expected to give rise to a shallow decay $t^{-1/3}$ or plateau
$t^{0}$, in the optical thin ($\nu_a<\nu_R$) or thick ($\nu_a>\nu_R$)
regimes, respectively. Another recognizable radio signature of a
jet-like geometry is the ``peak flux cascade'', in which successively
smaller frequencies reach lower peak fluxes (i.e. $F_m\propto
\nu_m^{1/2}$). Taken together, these observational signatures can be
used to infer the opening angles $\theta_j$ of wide angle jets. Such
jets are hard to detect at optical wavelengths because the break is
masked by the host galaxy, which typically dominates the light curve
between a week and a month after the burst \cite{bdf+01,fyb+03}.  Once
the real geometry of the outflow is known \cite{fks+01,bfk03} the
energy released in the GRB phase and the afterglow phase can be
determined.

As noted above, the radio band is fortuitously located close to
$\nu_a$ and as such it is a sensitive probe of the density structure
of the circumburst medium. Extensive broadband modeling \cite{pk02}
has yielded densities in the range 0.1 cm$^{-3} <\ $n$ <$\ 100
cm$^{-3}$, with a canonical value of order n$\simeq$10 cm$^{-3}$. Such
densities are found in the diffuse interstellar clouds of our Galaxy,
commonly associated with star-forming regions. A density of order 5-30
cm$^{-3}$ is also characteristic of the interclump medium of molecular
clouds, as inferred from observations of supernova remnants in our
Galaxy ({\em e.g.,} Chevalier 1999\nocite{chev99} and references
therein). Based on X-ray and optical observations alone, there have
been claims of high n$\gg10^{4}$ cm$^{-3}$ \cite{dl00,iz+01} or low
n$\ll 10^{-3}$ cm$^{-3}$ \cite{pk01} circumburst densities.  However,
in several of these cases when the radio data has been added to the
broadband modeling (i.e. constraining $\nu_a$), there is no longer any
support for either extreme of density \cite{hys+01,fyb+03}.

One unsolved problem on the structure of the circumburst environment
is the absence of an unambiguous signature of mass loss from the
presumed massive progenitor star in afterglow light curves
\cite{cl00}. Although there are some notable exceptions ({\em e.g.,}
Price et al.~2002\nocite{pbr+02}), most GRB light curves are best fit
by a jet expanding into a {\it constant} density medium instead of a
radial density gradient, $\rho\propto r^{-2}$ \cite{pk02}. Part of the
solution may lie in reduced mass loss rates due to metalicity effects,
or the motion of the star through a dense molecular cloud
\cite{wijers01}, both of which act to shrink the radius that the
pre-burst wind is freely expanding. It is equally likely that our
failure to distinguish between different models of the circumburst
medium is due to the lack of early afterglow flux measurements,
especially at millimeter and submillimeter wavelengths where the
largest differences arise \cite{pk00,yost03}. The resolution of this
conflict is important as it goes to the heart of the GRB progenitor
question.

At sufficiently late times, when the rest mass energy swept up by the
expanding shock becomes comparable to the initial kinetic energy of
the ejecta ($\sim$100 days), the expanding shock may slow to
non-relativistic speeds \cite{wrm97}. A change in the temporal slope is
expected at this time (Fig.~\ref{fig:cartoon}) with
$\alpha_{NR}=(21-15p)/10$ for a constant density medium, independent
of geometry. This dynamical transition provides a simple and power
method to derive the kinetic energy of the outflow which has expanded
to be quasi-spherical at this time. In contrast, most energy estimates
made at early times require knowledge of the {\it geometry} of the
outflow \cite{pk01,bkf03,bfk03}. Using the late-time radio light
curves and the robust Taylor-Sedov formulation for the dynamics we can
infer quantities such as the kinetic energy, ambient density, magnetic
field strength, and the size of the fireball. The radius can be
checked for consistency with the equipartition radius and the
interstellar scintillation radius. This method has been used for
GRB\,970508 \cite{fwk00} and for GRB\,980703 (Berger, {\it priv.
  comm.}), yielding energies of order $few\times 10^{50}$ erg, in
agreement with other estimates.

Finally, the radio light curves at late times may flatten due to the
presence of an underlying host galaxy. Most GRBs studied to date have
optical/NIR hosts but only about 20\% have been seen at centimeter and
submillimeter wavelengths \cite{bkf01,fbm+02,bck+03}. This radio
emission, if produced by star formation, implies star formation rates
SRF$\sim 500$ M$_\odot$~yr$^{-1}$ and L$_{bol}>10^{12}$ L$_\odot$,
clearly identifies these GRB hosts as ultraluminous starburst galaxies
which are all but obscured by dust at optical wavelengths.  This is an
emerging area with great potential for studying cosmic star formation
with a sample of galaxies selected quite differently than other
methods. Preliminary studies have already shown that GRB-selected
galaxies are significantly bluer than other radio-selected samples
\cite{bck+03}.

{\bf Acknowledgements.} DAF would like to thank his many collaborators
in the radio afterglow network, especially Shri Kulkarni and Edo
Berger.



\printindex

\begin{thebibliography}{10}

\bibitem{kp97}
Katz, J.~L. and Piran, T.
\newblock {\em ApJ}, {\bf 490}, 772, (1997).

\bibitem{goo97}
Goodman, J.
\newblock {\em New Astr.}, {\bf 2}(5), 449--460, (1997).

\bibitem{hbc+02}
{Hurley}, K., {\it et~al.}
\newblock {\em ApJ}, {\bf 567}, 447--453, (2002).

\bibitem{fkw+00}
{Frail}, D.~A., {\it et~al.}
\newblock in {\em AIP Conf. Proc. 526: Gamma-ray Bursts, 5th Huntsville
  Symposium},  298--302, (2000).

\bibitem{fkbw03}
{Frail}, D.~A., {\it et~al.} 
\newblock {\em AJ}, {\bf 125}, 2299--2306, (2003).

\bibitem{fjg+01}
{Fynbo}, J.~U., {\it et~al.} 
\newblock {\em A\&A}, {\bf 369}, 373--379, (2001).

\bibitem{bkb+02}
{Berger}, E., {\it et~al.} 
\newblock {\em ApJ}, {\bf 581}, 981--987, (2002).

\bibitem{pfg+02}
{Piro}, L., {\it et~al.} 
\newblock {\em ApJ}, {\bf 577}, 680--690, (2002).

\bibitem{dfk+01}
{Djorgovski}, S.~G., {\it et~al.} 
\newblock {\em ApJ}, {\bf 562}, 654--663, (2001).

\bibitem{fn85}
{Feigelson}, E.~D. and {Nelson}, P.~I.
\newblock {\em ApJ}, {\bf 293}, 192--206, (1985).

\bibitem{bkfs03}
Berger, E. {\it et~al.}
\newblock {A Radio Survey of Type Ib and Ic Supernovae: Searching for Engine
  Driven Supernovae}.
\newblock ApJ, in press; astro-ph/0307228, (2003).

\bibitem{kfw+98}
Kulkarni, S.~R., {\it et~al.} 
\newblock {\em Nature}, {\bf 395}, 663--669, (1998).

\bibitem{mes02}
{M{\' e}sz{\' a}ros}, P.
\newblock {\em Ann. Rev. Astr. Ap.}, {\bf 40}, 137--169, (2002).

\bibitem{gbb+00}
{Galama}, T.~J., {\it et~al.} 
\newblock {\em ApJ}, {\bf 541}, L45--L49, (2000).

\bibitem{fkn+97}
{Frail}, D.~A., {\it et~al.} 
\newblock {\em Nature}, {\bf 389}, 261--263, (1997).

\bibitem{rick90}
{Rickett}, B.~J.
\newblock {\em Ann. Rev. Astr. Ap.}, {\bf 28}, 561--605, (1990).

\bibitem{gfs+03}
{Galama}, T.~J., {\it et~al.} 
\newblock {\em ApJ}, {\bf 585}, 899--907, (2003).

\bibitem{wkf98}
{Waxman}, E., {Kulkarni}, S.~R., and {Frail}, D.~A.
\newblock {\em ApJ}, {\bf 497}, 288--293, (1998).

\bibitem{fwk00}
{Frail}, D.~A., {Waxman}, E., and {Kulkarni}, S.~R.
\newblock {\em ApJ}, {\bf 537}, 191--204, (2000).

\bibitem{kfs+99}
{Kulkarni}, S.~R., {\it et~al.} 
\newblock {\em ApJ}, {\bf 522}, L97--L100, (1999).

\bibitem{fbg+00}
{Frail}, D.~A., {\it et~al.} 
\newblock {\em ApJ}, {\bf 538}, L129--L132, (2000).

\bibitem{hys+01}
{Harrison}, F.~A., {\it et~al.} 
\newblock {\em ApJ}, {\bf 559}, 123--130, (2001).

\bibitem{yfh+02}
{Yost}, S.~A., {\it et~al.} 
\newblock {\em ApJ}, {\bf 577}, 155--163, (2002).

\bibitem{bsfk03}
{Berger}, E., {\it et~al.} 
\newblock {\em ApJ}, {\bf 587}, L5--L8, (2003).

\bibitem{abb+99}
{Akerlof}, C., {\it et~al.} 
\newblock {\em Nature}, {\bf 398}, 400--402, (1999).

\bibitem{sp99b}
{Sari}, R. and {Piran}, T.
\newblock {\em ApJ}, {\bf 517}, L109--L112, (1999).

\bibitem{sr02}
{Soderberg and Ramirez-Ruiz}.
\newblock Flaring up: Radio Diagnostics of the Kinematic, Hydrodynamic and
  Environmental Properties of GRBs.
\newblock MNRAS, in press; astro-ph/astro-ph/0210524, (2002).

\bibitem{hbf+99}
{Harrison}, F.~A., {\it et~al.} 
\newblock {\em ApJ}, {\bf 523}, L121--L124, (1999).

\bibitem{bsf+00}
{Berger}, E., {\it et~al.} 
\newblock {\em ApJ}, {\bf 545}, 56--62, (2000).

\bibitem{rho99}
{Rhoads}, J.~E.
\newblock {\em ApJ}, {\bf 525}, 737--749, (1999).

\bibitem{sph99}
{Sari}, R., {Piran}, T., and {Halpern}, J.~P.
\newblock {\em ApJ}, {\bf 519}, L17--L20, (1999).

\bibitem{bdf+01}
{Berger}, E., {\it et~al.} 
\newblock {\em ApJ}, {\bf 556}, 556--561, (2001).

\bibitem{fyb+03}
{Frail}, D.~A., {\it et~al.} 
\newblock {\em ApJ}, {\bf 590}, 992--998, (2003).

\bibitem{fks+01}
{Frail}, D.~A., {\it et~al.} 
\newblock {\em ApJ}, {\bf 562}, L55--L58, (2001).

\bibitem{bfk03}
Bloom, J.~S., Frail, D.~A., and Kulkarni, S.~R.
\newblock GRB Energetics and the GRB Hubble Diagram: Promises and Limitations.
\newblock ApJ in press, astro-ph/0302210, (2003).

\bibitem{pk02}
{Panaitescu}, A. and {Kumar}, P.
\newblock {\em ApJ}, {\bf 571}, 779--789, (2002).

\bibitem{chev99}
{Chevalier}, R.~A.
\newblock {\em ApJ}, {\bf 511}, 798--811, (1999).

\bibitem{dl00}
{Dai}, Z.~G. and {Lu}, T.
\newblock {\em ApJ}, {\bf 537}, 803--809, (2000).

\bibitem{iz+01}
{in' t Zand}, J., {\it et~al.} 
\newblock {\em ApJ}, {\bf 559}, 710--715, (2001).

\bibitem{pk01}
{Panaitescu}, A. and {Kumar}, P.
\newblock {\em ApJ}, {\bf 554}, 667--677, (2001).

\bibitem{cl00}
{Chevalier}, R.~A. and {Li}, Z.
\newblock {\em ApJ}, {\bf 536}, 195--212, (2000).

\bibitem{pbr+02}
{Price}, P.~A., {\it et~al.} 
\newblock {\em ApJ}, {\bf 572}, L51--L55, (2002).

\bibitem{wijers01}
{Wijers}, R.~A.~M.~J.
\newblock in {\em Gamma-ray Bursts in the Afterglow Era},  306--311, (2001).

\bibitem{pk00}
{Panaitescu}, A. and {Kumar}, P.
\newblock {\em ApJ}, {\bf 543}, 66--76, (2000).

\bibitem{yost03}
Yost, S. {\it et~al.}
\newblock {A Study of the Afterglows of Four GRBs: Constraining the Explosion
  and Fireball Model}.
\newblock ApJ, in press; astro-ph/0307056, (2003).

\bibitem{wrm97}
Wijers, R. A. M.~J., Rees, M.~J., and M\'esz\'aros, P.
\newblock {\em MNRAS}, {\bf 288}, L51--L56, (1997).

\bibitem{bkf03}
{Berger}, E., {Kulkarni}, S.~R., and {Frail}, D.~A.
\newblock {\em ApJ}, {\bf 590}, 379--385, (2003).

\bibitem{bkf01}
Berger, E., Kulkarni, S., and Frail, D.~A.
\newblock {\em ApJ}, {\bf 560}, 652--658, (2001).

\bibitem{fbm+02}
{Frail}, D.~A., {\it et~al.} 
\newblock {\em ApJ}, {\bf 565}, 829--835, (2002).

\bibitem{bck+03}
{Berger}, E., {\it et~al.} 
\newblock {\em ApJ}, {\bf 588}, 99--112, (2003).

\end{thebibliography}
\end{document}